\documentclass[preprint,12pt]{elsarticle}

\usepackage{graphicx}
\usepackage{amssymb}
\usepackage{color}

\usepackage{lineno}

\begin{document}

\begin{frontmatter}


\title{Kirchhoff's metasurfaces}



\author{Yoshiaki Nishijima}
\address{Department of Physics, Electrical and Computer Engineering, Graduate School of Engineering, Yokohama National University, 79-5 Tokiwadai, Hodogaya-ku, Yokohama 240-8501, Japan.}

\author{Armandas Bal\v{c}ytis}
\address{Centre for Micro-Photonics, Faculty of Science, Engineering and Technology, Swinburne University of Technology, Hawthorn, VIC 3122, Australia}
\address{Center for Physical Sciences and Technology, A. Go\v{s}tauto 9, LT-01108 Vilnius, Lithuania.}

\author{Shin Naganuma}
\address{Department of Physics, Electrical and Computer Engineering, Graduate School of Engineering, Yokohama National University, 79-5 Tokiwadai, Hodogaya-ku, Yokohama 240-8501, Japan.}

\author{Gediminas Seniutinas}
\address{Centre for Micro-Photonics, Faculty of Science, Engineering and Technology, Swinburne University of Technology, Hawthorn, VIC 3122, Australia}
\address{Current address,  Paul Scherrer Institute, Villigen CH-5232, Switzerland}

\author{Saulius Juodkazis}
\address{Centre for Micro-Photonics, Faculty of Science, Engineering and Technology, Swinburne University of Technology, Hawthorn, VIC 3122, Australia}

\begin{abstract}
Thermo-optical properties of the nanodisc and metal hole array plasmonic perfect absorber (PPA) metasurfaces were designed and characterised at mid-infrared wavelengths. Both, light emitter and detector
systems are highly thought after for the future sensor networks in the internet-of-things for various spectral domains. Reciprocity of the absorbance and emittance is shown experimentally, i.e., the PPAs are following Kirchhoff's law where the patterns exhibiting a strong optical absorption were
found enhanced thermal radiation. Design principles and scaling for photo-thermal conversion are discussed. The highest efficiency of light-to-heat and
heat-to-radiation were obtained for the Au-Si-Au structures.
\end{abstract}


\end{frontmatter}



Control of thermal radiation is one of challenges in the future's essential technologies, which could deliver light emitters with tailored spectral, polarization, directionality properties strongly required for the infrared (IR) spectral window. According to the Einstein's description of emission, the  coefficient, $A$, for the probability of spontaneous optical
emission is smaller (a lower optical transition probability) at the mid-IR wavelength region as compared with the visible spectral range. Therefore, it would be expected that a mid-IR photo diode would have a low quantum efficiency of radiation/emission. For mid-IR wavelengths, the molecular vibrations have the representative absorbance bands and are used for
qualitative and quantitative detection of different compounds,
hence, realizing a molecular finger printing. Especially the mid-IR
technology has focused on the gas sensing, exhaust control in
cars, environmental monitoring, and human health care. The cavity
ring down spectroscopy where a high finesse resonator is coupled with quantum cascade laser (QCLs) or a laser comb is one of the successful and promising
technologies for optically detection of low concentration
molecular species~\cite{CRDS1, CRDS2, CRDS3, CRDS4, CRDS5}. However, the cost of equipment such as QCL and
comb lasers, highly reflectivity mirrors ($R > 99\%$) at specific wavelengths, high speed photodiode detectors are still making this technology expensive
and less portable.

\begin{figure}[tb]\begin{center}
\includegraphics[width=12.0cm]{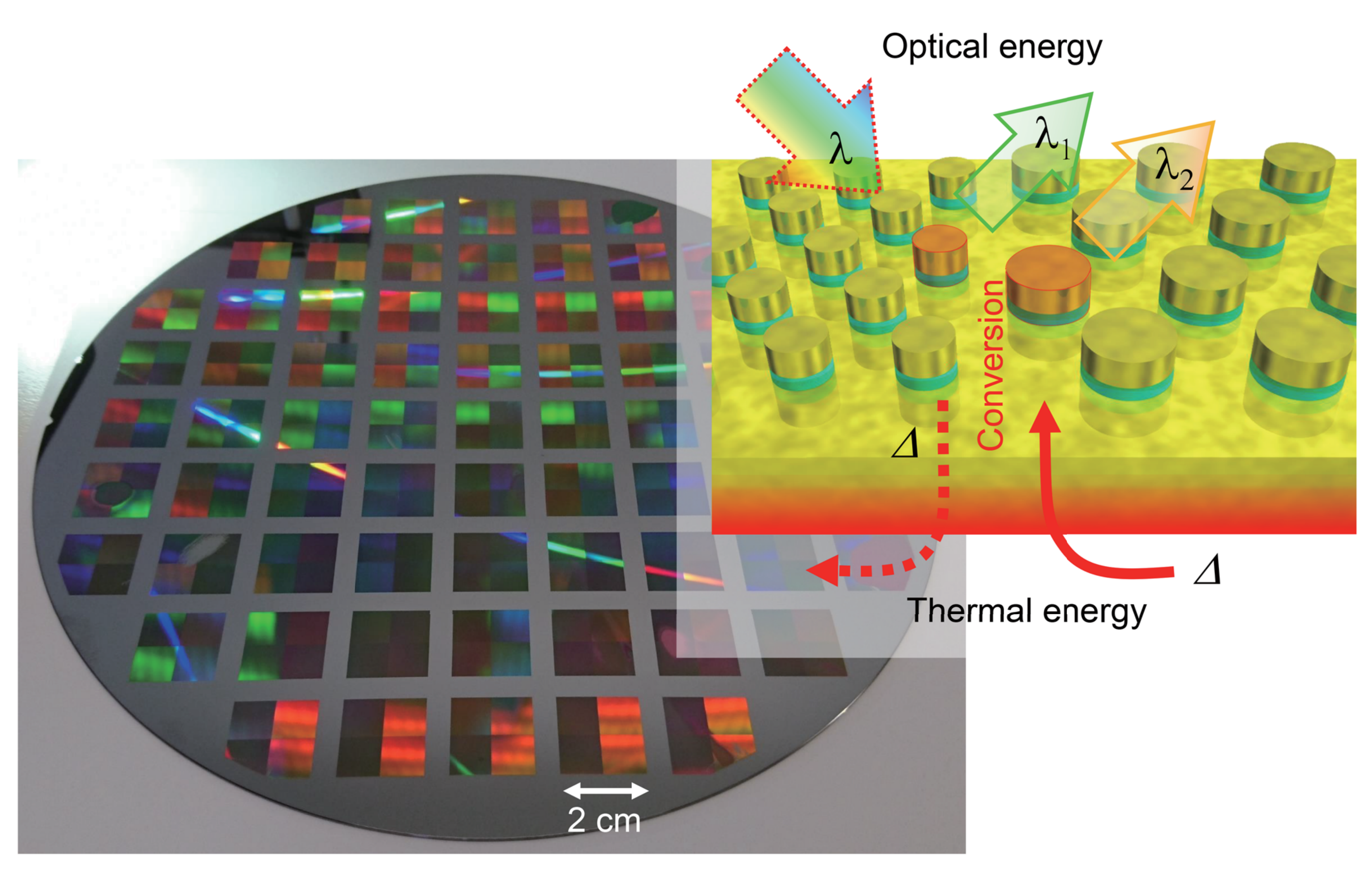}
\caption{Concept of the Kirchhoff's metasurface devices fabricated by mass production photo-lithography on 8-inch Si wafers. Illustration of optical-to-thermal and thermal-to-optical energy conversions by the Kirchhoff's metasurfaces; $\Delta$ marks the converted portion of energy.}
\label{f0} \end{center}
\end{figure}

Nowadays, the miniaturisation trend and information technology
requires sensors for realizing the Internet-of-things (IoT) to
become the founding principle of the future technologies~\cite{IoT1,IoT2,IoT3,IoT4,IoT5}. The demand of IoT projects the following directions for sensors: low cost, portable (wearable), long lifetime (low energy consumption), and large quantities of sensors interconnected in the network~\cite{enose1,enose2,enose3}. 
Plasmonics and nanophotonics are
the powerful candidates to contribute and solve the above
mentioned challenges for IoT, especially for the chemical sensing
field. An enhancement of the IR absorption by plasmons on nano-textured/pattered surfaces utilized in
the surface-enhanced IR absorption (SEIRA) spectroscopy realize
the high sensitivity and scaling along with the cost-down trend for entire sensing systems~\cite{nishijima1,nishijima2,nishijima3,halas1,halas2}. 
When the light emitters and detectors consist of the plasmonic materials, the strong EM-field enhancement factors would be expected for the enhanced emission efficiency as well as for the light absorption (the Kirchhoff's law).
A practical design of plasmon perfect absorbers (PPAs) has a configuration of a metal film, insulator (SiO$_2$, Si, TiO$_2$ or other dielectric), and
metal nanostructure. Optical properties of the PPAs are zero transmission, $T=0$, and close to zero reflection, $R = 0$, at the plasmon resonance. At the wavelength of plasmonic resonance, light should be
absorbed by PPA~\cite{diem,miyazaki1,miyazaki2,miyazaki3,takahara1,takahara2,maruyama,ppasensor,ppareview,sergey,oliver}.
For the ideal case, 100\% of incident light can be absorbed at the resonance wavelength by PPA.

Kirchhoff's law of radiation stipulates that the strong light
absorber is also an efficient light emitter. Therefore PPA
structures are expected to be also applicable for the light
emitters, i.e., thermo-optical input/output devices. When plasmon
materials absorbs at a specific wavelength, a decaying oscillation
according to the quality Q-factor of the plasmonic resonance
- oscillation of free electrons - generates Joule heat which is
radiated. Experimental verification of Kirchhoff's law at mid-far-IR wavelengths and for different absorption mechanisms in nanophotonic patterns and structures is strongly required. 

Here, we show a design of PPAs at IR wavelengths and tested them as
optical emitters at IR wavelength. We experimentally characterise thermo-optical input and output properties of PPAs and show the reciprocity correlation between the absorbance and emittance of these Kirchhoff's metasurfaces as shown in Fig.\ref{f0}.

\begin{figure}[tb]
\begin{center}
\includegraphics[width=13cm]{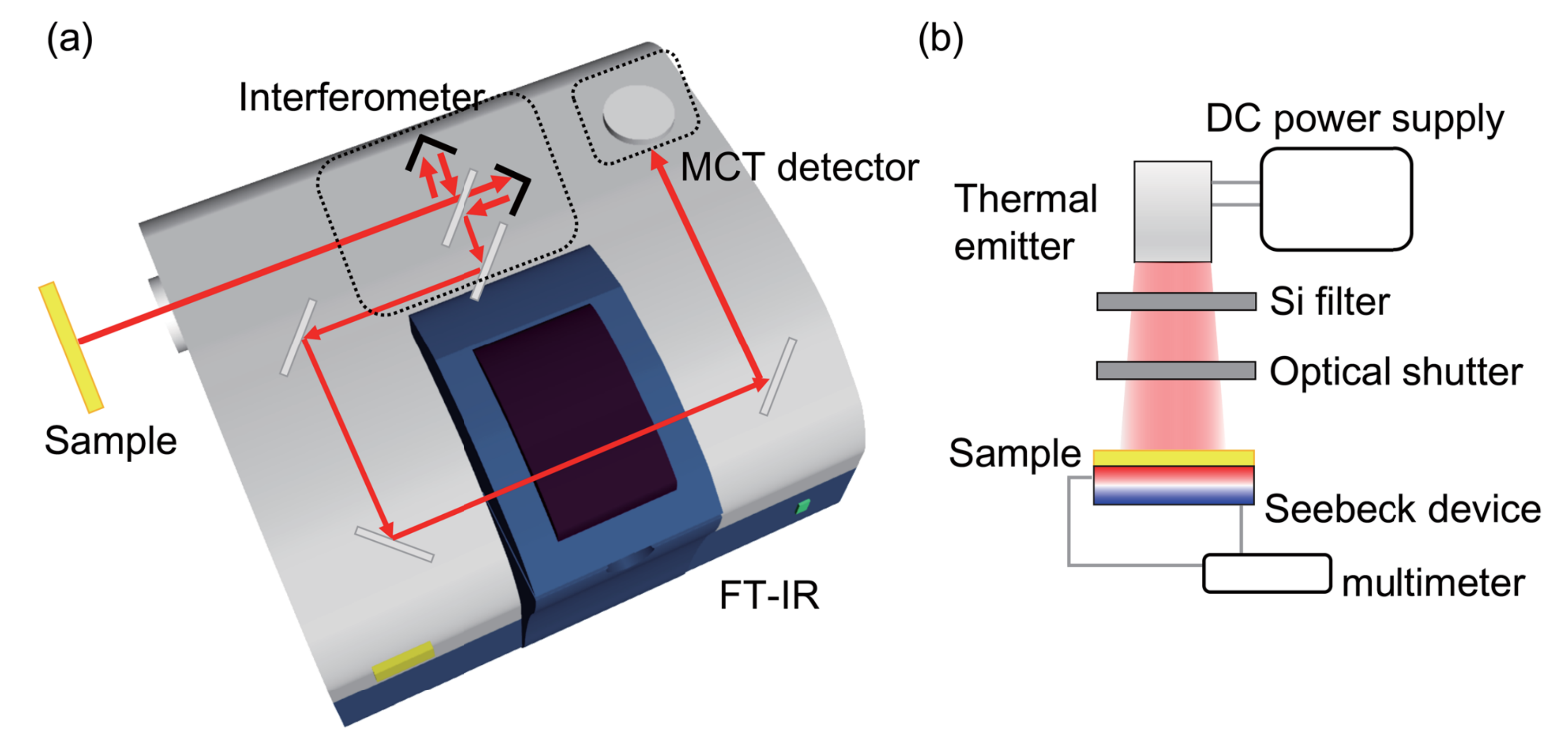}
\caption{Schematic illustration of the optical measurement setup.
(a) Principle of a FT-IR measurement of thermal radiation; MCT is the 
mercury cadmium telluride IR detector. (b) Setup for the
photo-thermal conversion. A broadband tungsten thermal emitter with Si window to cut visible emission was used as a thermal light source. } \label{f1} \end{center}
\end{figure}

\section{Experimental}

\subsection*{Fabrication of large area PPAs}

Two type metasurfaces PPAs, the nano-disk and multi-hole array
(NDA and MHA) nanostructures were fabricated on a silicon wafer
via reduction projection photolithography  using an $i$-line
stepper (NSR205-i14E, NIKON Co.)~\cite{nishijima4,PPAs}. 
First, a 200-nm-thick Au film has sputtered on the top of a double side polished Si wafer. The same photo mask pattern was used for both NDA and MHA fabrication.
For NDAs a positive-tone photo-resist (TLOR-P003 HP, Tokyo Ohka
Kogyo Co.), whereas for the MHA structures the negative
tone resist (TLOR-N001 PM, Tokyo Ohka Kogyo Co.) was used. The
resist pattern was developed for the following deposition of the insulator and metal structures deposited by magnetron sputtering with a sequence of 10~nm of
Si/SiO$_2$ and 50~nm of Au with a 3~nm Ti adhesion layer between
Au and insulator (AXXIS, JKLesker). Then, the lift-off process followed in acetone for NDAs or methyl isobutyl ketone heated on a hot plate for MHAs.
Both structures were rinsed by isopropanol and dried under nitrogen stream.

\begin{figure}[tb]
\begin{center}
\includegraphics[width=14.5cm]{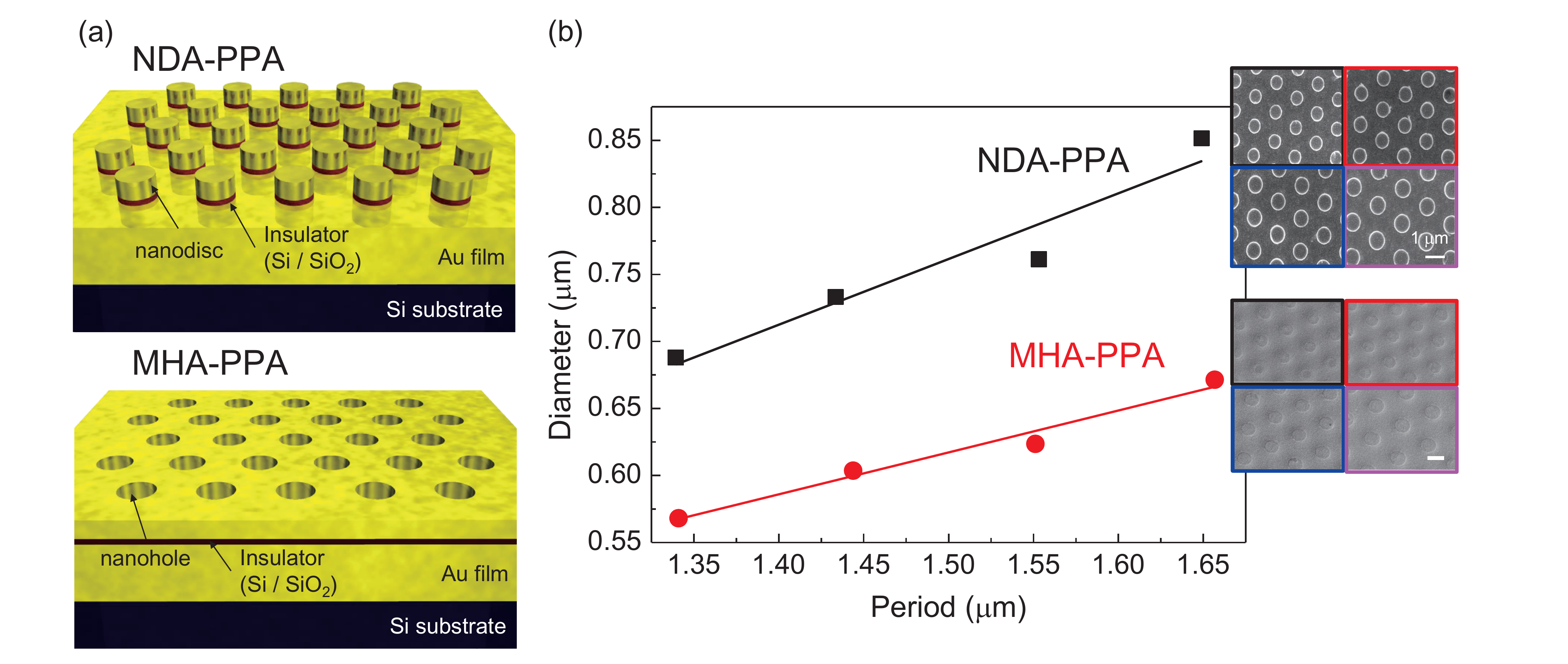}
\caption{(a) Schematic illustrations of PPAs structures of NDA and MHA types. (b) Disc (hole) diameter vs. period of the fabricated
NDA and MHA structures.} \label{f2}
\end{center}
\end{figure}

\subsection*{Optical characterization of PPAs}

The optical reflection spectra were measured by conventional
technique that is a combination of the FT-IR (FT-IR 4200, JASCO
Co.) with a microscope unit (IRT-1000). As reference reflectivity,
the Au mirror with 98\% absolute reflectance was used. For the
thermal radiation spectroscopy, the 8$^\times$ times magnification
Cassegrain reflector lens with numerical aperture $NA=0.5$ was used. By controlling a square aperture, the 500 $\times$ 500~$\mu$m$^2$ region was measured.

\subsection*{Thermal radiation and photo-thermal conversion}

The thermal radiation and photo-thermal generation were
measured by a custom home built setup. For realisation of FT-IR spectroscopy setup, the emitted light was passed through an interferometer. We have
customized the commercial FT-IR (FT-IR 4200, JASCO Co.)
as shown in Fig.~\ref{f1}(a). Light was coupled from outside into
the FT-IR setup via an optical side-port. It passed through the
interferometer and was detected on the HgCdTe (MCT) detector.
Sample chamber was evacuated by oil rotary pump to reduce the
absorption effect by CO$_2$, H$_2$O and other atmospheric gases in the
present inside the measurement chamber. The target substrate was contacted to the ceramic heaters on the Al plate. 
The sample was heated at 300$^{\circ}$C; the temperature was monitored using radiative infrared thermometer.

Materials radiate the black body radiation according to their temperature with part of emission at the IR wavelengths. Most of the metals, which have high
reflectivity have small emissivity. Therefore, an Al plate was
used to block the IR thermal radiation that was generated from
surrounding objects  heated to 300$^{\circ}$C by the heater.
Ceramic heater was also covered by Al to screen thermal radiation.
To reduce the thermal noise form other sources, a top cover plate
made by Al with 1$\times$1~cm$^2$ square through-hole was used and
all the system was covered with Al film. This was required to
measure the radiation from the sample. The radiation efficiency
was determined using a relative radiation intensity of 94\% of the
black body using black ink (THL-1B TASCO Co.).

Photo-thermal conversion was measured using the Seebeck device as
shown in Fig.~\ref{f1}-(b). Tungsten filament-based thermal
radiation emitter was used for the IR-light source. A
pseudo-collimated light was illuminated onto samples through a Si
wafer, which works as a cut-filter under the 11its 00~n upgth
light. The Si filter was separated enough from the heater to prevent
its heat up by the tungsten heater. Sample was put on the Seebeck
device on the Al plate. For the photo-thermal conversion, the
sensitivity of Seebeck device was low, therefore a larger 2$\times$2~cm$^2$
region has illuminated, which contains four regions which were adjacent to each other. The output voltage from the Seebeck device has measured with a multimeter.

\section{Results and discussion}

Figure~\ref{f2}(a) shows a schematic illustration of both, the nano-disk array (NDA) and multi-hole array (MHA) structures used for PPAs. Here, both metals in the PPA were gold because of the chemical stability for high temperature
cycling in air conditions. As the insulator layer, Si and SiO$_2$
were used to test influence of strongly different refractive index materials. The PPA patterns were fabricated uniformly over the 1$\times$1~cm$^2$ area.
Different hole (disk) diameters and periods of the pattern were
tested. The periods and diameters of the NDA and MHA structures is plotted in
the Fig.~\ref{f2}(b). The same mask was used for both, the NDA and
MHA structures and the period was reliably reproduced. However,
the diameters of the disks and holes were slightly different. The
ratio of the diameter-to-period was set to the 1:2. The NDA
structures were reproduced according to the design, while the MHA
patterns showed slightly smaller holes. This is due to different retention
of the polymer vs. dose for the used negative and positive tone
resists.

\begin{figure}[tb]
\begin{center}
\includegraphics[width=12cm]{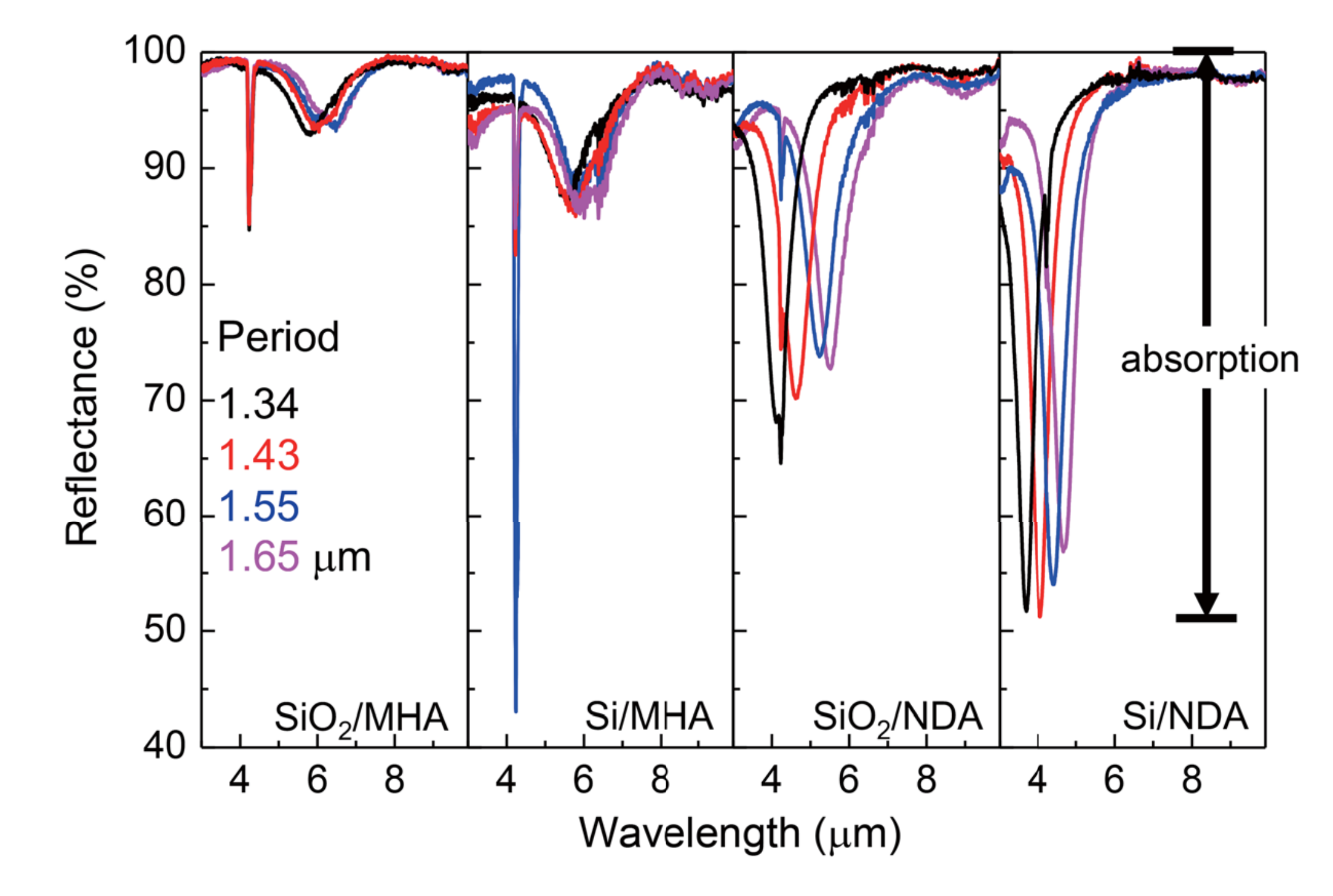}
\caption{Optical reflection spectra of PPAs with NDA and MHA
patterns of different periods and with Si and SiO$_2$ insulators.
The CO$_2$ absorption at 4.3~$\mu$m shows a sharp spectral
artefact.} \label{f3}
\end{center}
\end{figure}

Figure~\ref{f3} shows the optical reflection spectra for the NDA
and MHA structures. The PPAs had a 200-nm-thick Au film on the
bottom of the structure. Therefore no transmission of light can be
observed in experiment (transmission data has been omitted in Fig.\ref{f3}) and light (except scattering and reflection) is absorbed
in the PPA structure.
The sharp peaks around 4.3~$\mu$m are caused by CO$_2$ absorption in air, which could be removed by N$_2$ purging when required.

Noteworthy, when we consider typical plasmonic resonances,
the NDAs were inspected in the reflection or scattering while the MHAs were measured in
the transmission mode for comparison of their performance as
absorbers. The MHA showed considerably smaller
reflection dip, which means a weaker absorption at the plasmonic
resonance. However, the NDAs showed a strong absorption, hence,
modulation of the reflectance at the resonance wavelength. For both
structures, the Si insulator was preferable due to the larger
absorption as compared with SiO$_2$ (Fig.~\ref{f3}).

The plasmon peak wavelength is dependent on the geometry of the
structure. In NDAs, it is mainly defined by the diameter of the
disc, while in MHA by the period. Considering typical MHA
structures, their performance as absorbers had a weaker dependence
on the structure period (Fig.~\ref{f3}).

\begin{figure}[tb]
\begin{center}
\includegraphics[width=12cm]{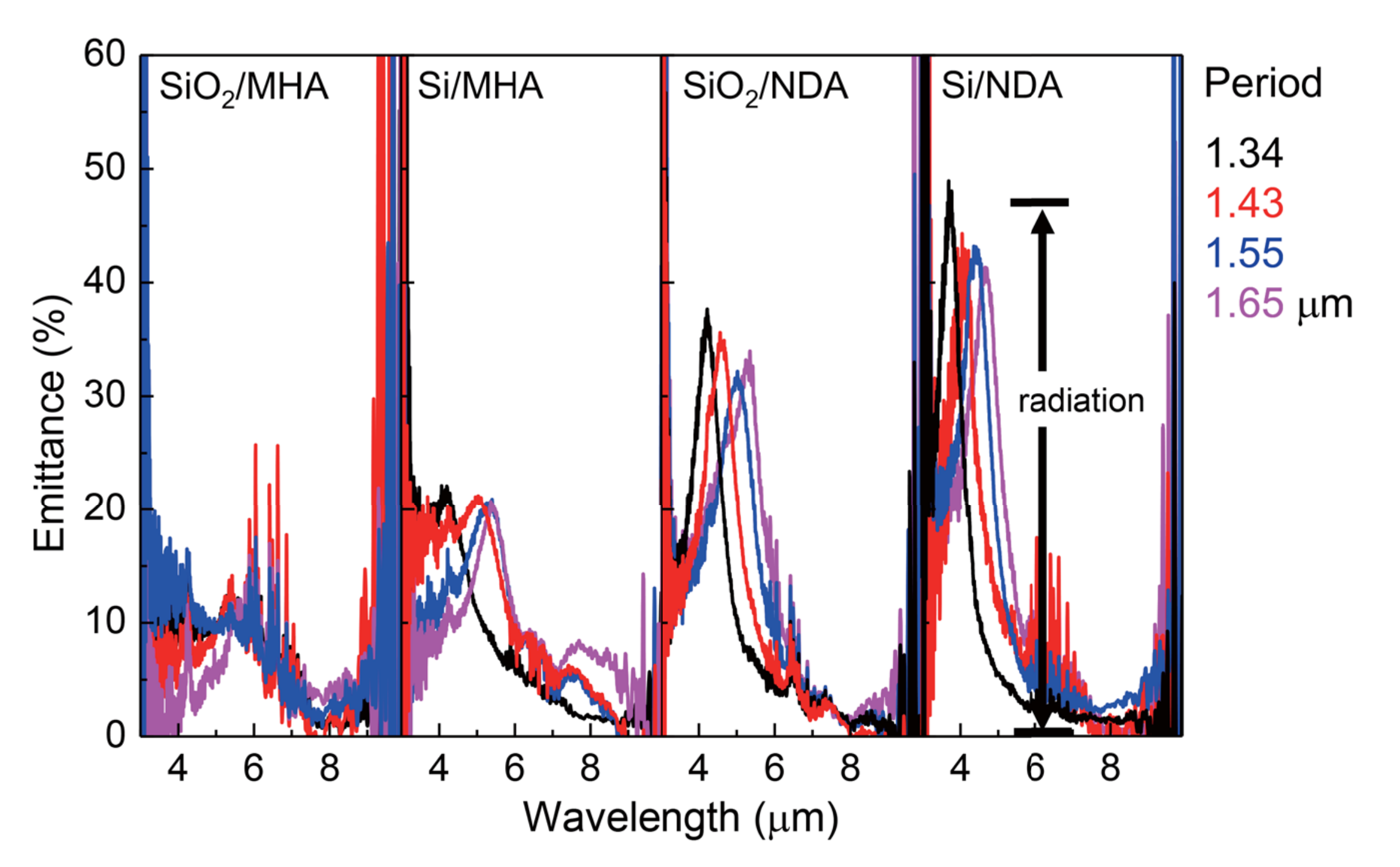}
\caption{Thermal radiation spectra of the NDA and MHA structures.} \label{f4}
\end{center}
\end{figure}

Figure~\ref{f4} summaries thermal radiation properties of the
very same PPAs characterized above for their absorbance. To obtain
emittance, the radiation spectra was normalized to the sample that
was covered with a 94\% radiation efficiency black body ink. As
expected from the Kirchhoff's law, the spectral shape of emittance
was strongly anti-correlated with the reflectivity spectrum of the same
structure (or the absorbance is correlated with emittance). The peak emittance agreed well with the spectral position
of absorption maximum (a dip in reflectance). According to the
Kirchhoff's law of the thermal radiation, the thermal emission and
absorption are expected to be equal under the thermodynamic
equilibrium condition.

The NDA absorber structure with Si insulator presents the most
effective emitter among the tested four. With the PPA it would be
possible to realize the perfect absorption: a zero-transmission and
reflection, hence, a 100\% absorption, when the impedance matching to
the air has been achieved~\cite{ppa}. For this purpose, the
dielectric permittivity of the structures, substrate and insulator
layer and their thickness optimization is required.

\begin{figure}[tb]\begin{center}
\includegraphics[width=14.5cm]{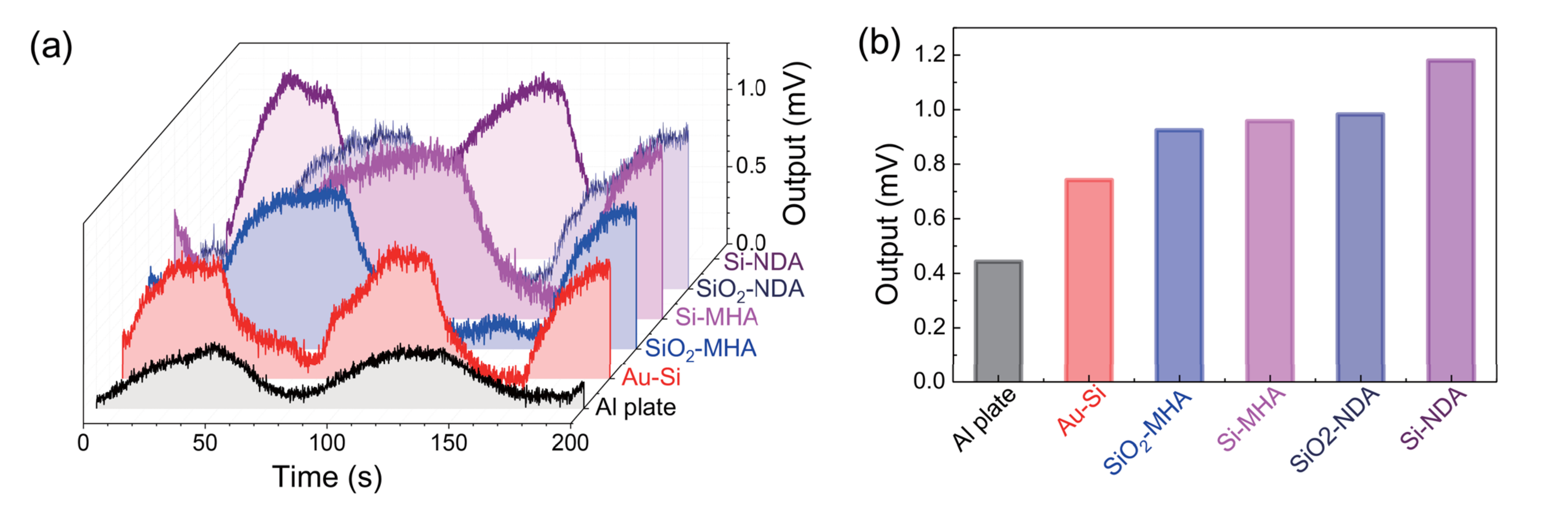}
\caption{(a) Time dependence of the output voltage from Seebeck
device with NDA/MHA absorbers. For comparison, 200~nm Au deposited
on Si substrate (Au-Si) and 0.5~mm Al plate were measured. (b) The
maximum (saturation) output voltage for each substrates.}
\label{f5} \end{center}
\end{figure}

Thermo-emitter response of the PPAs is shown in Fig.~\ref{f5} where Seebeck voltage is presented (see, Fig.~\ref{f1}(b)).
A time dependent response to the on-off
switching of the IR light is presented in (a); switching was controlled by a mechanical shutter within 0.7~s time intervals. Both "on" and "off" response times to the
saturated state of the Seebeck voltage took few tens of seconds for all samples. This is caused by the time required to diffuse heat through an entire
thickness of the substrate. In Fig.~\ref{f5}(b), the maximum
output voltage is shown. All the PPA structures exhibited a larger
efficiency of thermal emission as compared to the 0.5-mm-thick Al
plate and 200-nm-thick Au films on Si (which is the same substrate
without top nanostructures). Metals such as Al and Au have a lower
absorption coefficient and consequently have a lower emissivity:
Al 4-8\% and  Au 2-3\%, respectively~\cite{rad}. Even in the mid-IR range, there is a weak interband absorption in metals (hence emission). Both metal surface have some roughness that can facilitate coupling of the incident light into a surface plasmon polariton wave. Therefore, there were still some
photo-thermal heat generation  present using these substrates. It
is important to separate the effect which is not caused by the
designed structure, especially for the Au-Si substrate with the
same composition as PPAs. The output form the Au-Si served as a
reference for the background of the plasmonic absorber/emitter (Fig.~\ref{f5}(b)).

Experiments clearly show that the thermal emission from PPAs was
stronger than background noise from unstructured samples made from
the same materials. The Seebeck device (Fig~\ref{f1}(b)) used in this study was linear at around room temperature~\cite{komatsu}. Thermal emission
from PPAs was estimated to be at least from 2.5 to 3.0 times
stronger than that from the Al plate.

The tungsten thermal emitter (Fig.~\ref{f1}(b)) covers a wide wavelength range where it performs close to the blackbody radiation efficiency. In this
experiment, samples with four different periods were illuminated.
The plasmon band of the PPAs are narrow in the tested wavelength
range and the differences in geometry between samples were rather
small. The averaged PPA absorbance was from 2.7 to 7.8\% in this
wavelength region. Therefore the output thermal emission was
expected to be small. Illumination of PPAs by monochromatic light
from mid-IR laser, light emitting diode can be used to
characterise PPA detectors and is considered for a future study.

The temperature change was estimated to be from 0.01~K (Al
plate) to 0.03~K (Si-NDA). A large heat conductivity and thick
substrate would dissipate heat to air and it decrease contrast
from the thermal background. The response time to reach saturation
(equilibration) is dependent of the thermal capacitance. Integration
of PPAs on thermally isolated membrane could help to reduce the
thermal background noise and to increase the temperature of the
PPA. It could help to realize a plasmon micro-bolometer working at
room temperature.

\begin{figure}[tb] \begin{center}
\includegraphics[width=14.5cm]{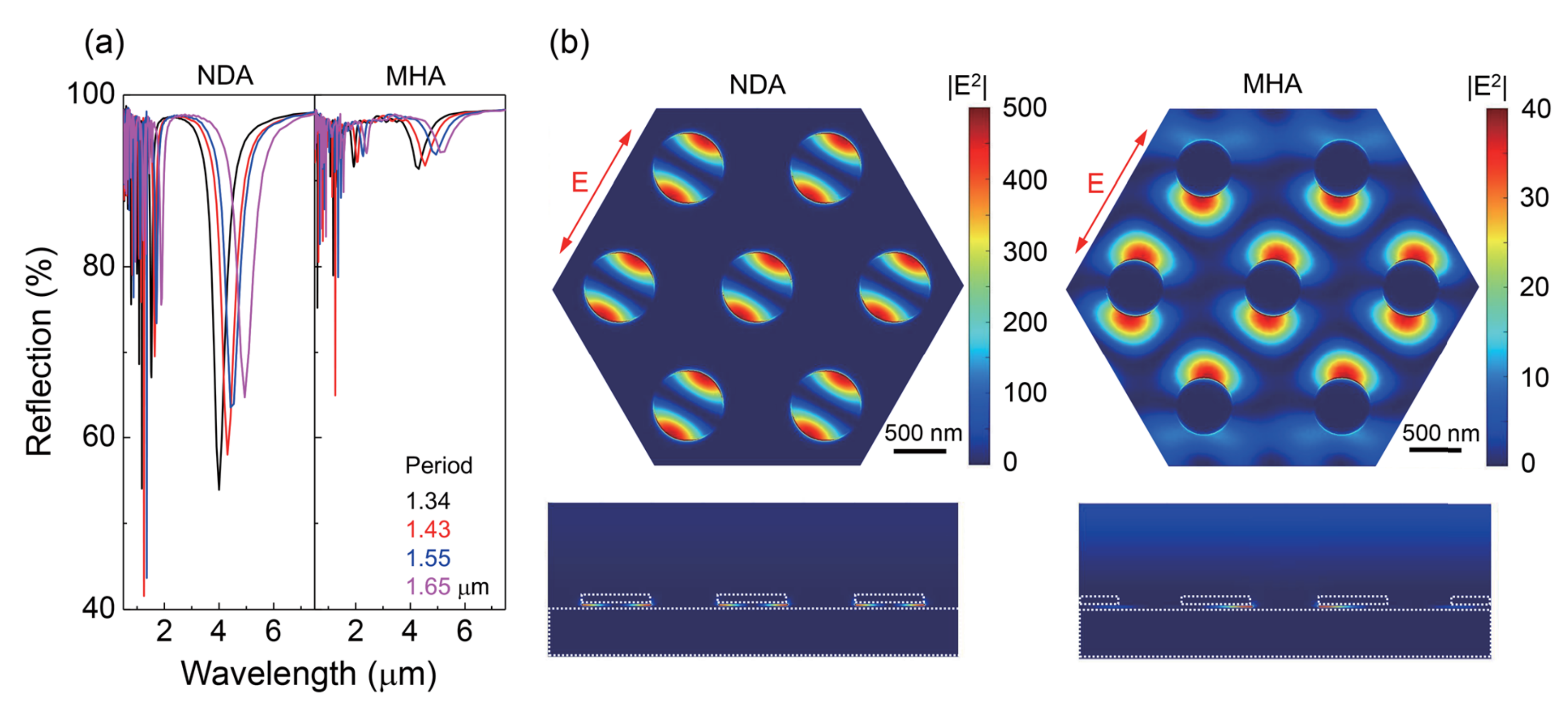}
\caption{FDTD simulation for SiO$_2$ NDA and MHA structures.
Reflection spectra (a) and electromagnetic field profile (b) of both structures at resonance. Electrical E-field was linearly polarised as marked by arrows.} \label{f6}\end{center}
\end{figure}

\subsection*{FDTD calculations}

The FDTD calculations were performed for the ideal shape of the
flat surface of the metal and insulator. Figure~\ref{f6}(a) shows the optical reflection spectra of PPAs as
used in experiments. Absorbance of NDA was more than 4 times
larger than that of MHA. From the electromagnetic field profile at
the resonance wavelength shown in Fig.~\ref{f6}(b) it follows that the E-field is localized in the insulator layer. As described in the experimental
section, the insulator layer was at the bottom of metal
structures. However in the case of MHA-PPAs, the insulator layer
was continuous overall the entire sample, E-field is delocalized
and absorption was smaller than that of NDA structures.

The basic scaling behaviour of PPAs observed in experiments and differences between MHA and NDA patterns can be gauged from this simple FDTD model to reveal the localization of energy and local field enhancement. In the case of SiO$_2$ insulator layer, the modeled optical properties were closely following the experimental results. 

As our previous work indicated, the surface roughness of the structures, especially the metal base, affects numerical results considerably. Therefore it is important to reproduce a closely matching resemblance of the
experimental PPA when it is rendered for FDTD modeling seeking more quantitative match with experiments. It was found that FDTD predictions of PPAs with Si insulator layer show differences with experimental observations due to electrical conductivity of Si (which is accounted for via 
permittivity of materials used in numerical simulations)~\cite{PPAs}.

\section*{Conclusions and Outlook}

It is demonstrated that NDA and MHA structures made for resonant
absorbance are performing as the narrow band thermal radiation emitters. The energy conversion for thermal energy harvesting as well as extraction into
radiation are feasible. The most efficient thermal emitters had the NDA design with Si as insulator for the IR wavelengths.

Kirchhoff's metasurfaces can find applications for thermal
emitters and detectors and open a toolbox for engineering thermal
energy control for miniaturised (wavelength scale) devices in IR spectral
range. Since Si has comparatively high transparency in THz
spectral range, we can envisage application of such Kirchhoff's
metasurfaces over a broad spectral range. Next modality in IR emission control is creation of directional IR emitters by harnessing the Wolf's effect~\cite{nature} with a grating over the Kirchhoff's metasurfaces. Such surface will deliver angular selectivity and enhancement of light extracted with the sub-wavelength grating realizing a coherent IR emitter~\cite{nature,liu}.   
\section*{acknowledgement}
 Authors are grateful for the Tokyo Ohka Kogyo Co. Ltd. for access to stepper. YN and SN gratefully thank Dr. Hideki Miyazaki from National institute for
material science for advise regarding construction of thermal
radiation measurement setup. YN is grateful for partial support by
Japan Society for the Promotion of Science (JSPS), Grants-in-Aid
for Scientific Research,  Open Partnership Joint Projects of JSPS Bilateral Joint Research Projects, Tateishi and Amada fundations.
SJ acknowledges partial support via the Australian Research
Council Discovery project DP130101205 and a startup funding of
Nanotechnology facility by Swinburne University.

\bibliographystyle{model1-num-names}
\bibliography{kirchhoff.bib}

\begin{thebibliography}{36}
\expandafter\ifx\csname natexlab\endcsname\relax\def\natexlab#1{#1}\fi
\providecommand{\bibinfo}[2]{#2}
\ifx\xfnm\relax \def\xfnm[#1]{\unskip,\space#1}\fi
\bibitem[{Wheeler et~al.(1998)Wheeler, Newman, Orr-Ewing, and Ashfold}]{CRDS1}
\bibinfo{author}{M.~D. Wheeler}, \bibinfo{author}{S.~M. Newman},
  \bibinfo{author}{A.~J. Orr-Ewing}, \bibinfo{author}{M.~N.~R. Ashfold},
\newblock \bibinfo{title}{Cavity ring-down spectroscopy},
\newblock \bibinfo{journal}{J. Chem. Soc., Faraday Trans} \bibinfo{volume}{94}
  (\bibinfo{year}{1998}) \bibinfo{pages}{337--351}.
\bibitem[{Zalicki and Zare(1995)}]{CRDS2}
\bibinfo{author}{P.~Zalicki}, \bibinfo{author}{R.~N. Zare},
\newblock \bibinfo{title}{Cavity ring-down spectroscopy for quantitative
  absorption measurements},
\newblock \bibinfo{journal}{J. Chem. Phys.} \bibinfo{volume}{102}
  (\bibinfo{year}{1995}) \bibinfo{pages}{2708}.
\bibitem[{Berden et~al.(2000)Berden, Peeters, and Meijer}]{CRDS3}
\bibinfo{author}{G.~Berden}, \bibinfo{author}{R.~Peeters},
  \bibinfo{author}{G.~Meijer},
\newblock \bibinfo{title}{Cavity ring-down spectroscopy: Experimental schemes
  and applications},
\newblock \bibinfo{journal}{Int. Rev. Phys. Chem.} \bibinfo{volume}{19}
  (\bibinfo{year}{2000}) \bibinfo{pages}{565--607}.
\bibitem[{Sun et~al.(2018)Sun, Ding, Liu, Yang, and Li}]{CRDS4}
\bibinfo{author}{J.~Sun}, \bibinfo{author}{J.~Ding}, \bibinfo{author}{N.~Liu},
  \bibinfo{author}{G.~Yang}, \bibinfo{author}{J.~Li},
\newblock \bibinfo{title}{Detection of multiple chemicals based on external
  cavity quantum cascade laser spectroscopy},
\newblock \bibinfo{journal}{Spectroc. Acta Pt. A-Molec. Biomolec. Spectr.}
  \bibinfo{volume}{191} (\bibinfo{year}{2018}) \bibinfo{pages}{532--538}.
\bibitem[{Kiseleva et~al.(2018)Kiseleva, Mandon, Persijn, and Harren}]{CRDS5}
\bibinfo{author}{M.~Kiseleva}, \bibinfo{author}{J.~Mandon},
  \bibinfo{author}{S.~Persijn}, \bibinfo{author}{F.~Harren},
\newblock \bibinfo{title}{Line strength measurements and relative isotopic
  ratio 13c/12c measurements in carbon dioxide using cavity ring down
  spectroscopy},
\newblock \bibinfo{journal}{J. Quant. Spectrosc. Radiat. Transf.}
  \bibinfo{volume}{204} (\bibinfo{year}{2018}) \bibinfo{pages}{152--158}.
\bibitem[{Bandodkar et~al.(2016)Bandodkar, Jeerapan, and Wang}]{IoT1}
\bibinfo{author}{A.~J. Bandodkar}, \bibinfo{author}{I.~Jeerapan},
  \bibinfo{author}{J.~Wang},
\newblock \bibinfo{title}{Wearable chemical sensors: present challenges and
  future prospects},
\newblock \bibinfo{journal}{ACS Sens.} \bibinfo{volume}{1}
  (\bibinfo{year}{2016}) \bibinfo{pages}{464--482}.
\bibitem[{Khan et~al.(2015)Khan, Lorenzelli, and Dahiya}]{IoT2}
\bibinfo{author}{S.~Khan}, \bibinfo{author}{L.~Lorenzelli},
  \bibinfo{author}{R.~S. Dahiya},
\newblock \bibinfo{title}{Technologies for printing sensors and electronics
  over large flexible substrates: A review},
\newblock \bibinfo{journal}{IEEE Sens. J.} \bibinfo{volume}{15}
  (\bibinfo{year}{2015}) \bibinfo{pages}{3164--3185}.
\bibitem[{Caldara et~al.(2016)Caldara, Colleoni, Guido, Re, and Rosace}]{IoT3}
\bibinfo{author}{M.~Caldara}, \bibinfo{author}{C.~Colleoni},
  \bibinfo{author}{E.~Guido}, \bibinfo{author}{V.~Re},
  \bibinfo{author}{G.~Rosace},
\newblock \bibinfo{title}{Optical monitoring of sweat ph by a textile fabric
  wearable sensor based on covalently bonded
  litmus-3-glycidoxypropyltrimethoxysilane coating},
\newblock \bibinfo{journal}{Sens. Actuators, B} \bibinfo{volume}{222}
  (\bibinfo{year}{2016}) \bibinfo{pages}{213--220}.
\bibitem[{Senior(2014)}]{IoT4}
\bibinfo{author}{M.~Senior},
\newblock \bibinfo{title}{Novartis signs up for google smart lens},
\newblock \bibinfo{journal}{Nat. Biotechnol} \bibinfo{volume}{32}
  (\bibinfo{year}{2014}) \bibinfo{pages}{856}.
\bibitem[{Mannoor et~al.(2012)Mannoor, Tao, Clayton, Sengupta, Kaplan, Naik,
  Verma, Omenetto, and McAlpine}]{IoT5}
\bibinfo{author}{M.~S. Mannoor}, \bibinfo{author}{H.~Tao},
  \bibinfo{author}{J.~D. Clayton}, \bibinfo{author}{A.~Sengupta},
  \bibinfo{author}{D.~L. Kaplan}, \bibinfo{author}{R.~R. Naik},
  \bibinfo{author}{N.~Verma}, \bibinfo{author}{F.~G. Omenetto},
  \bibinfo{author}{M.~C. McAlpine},
\newblock \bibinfo{title}{Graphene-based wireless bacteria detection on tooth
  enamel},
\newblock \bibinfo{journal}{Nat. Commun} \bibinfo{volume}{3}
  (\bibinfo{year}{2012}) \bibinfo{pages}{763}.
\bibitem[{G\"{u}ntner et~al.(2016)G\"{u}ntner, Koren, Chikkadi, Righettoni, and
  Pratsinis}]{enose1}
\bibinfo{author}{A.~T. G\"{u}ntner}, \bibinfo{author}{V.~Koren},
  \bibinfo{author}{K.~Chikkadi}, \bibinfo{author}{M.~Righettoni},
  \bibinfo{author}{S.~E. Pratsinis},
\newblock \bibinfo{title}{E-nose sensing of low-ppb formaldehyde in gas
  mixtures at high relative humidity for breath screening of lung cancer},
\newblock \bibinfo{journal}{ACS Sens.} \bibinfo{volume}{1}
  (\bibinfo{year}{2016}) \bibinfo{pages}{528--535}.
\bibitem[{Maruyama et~al.(2018)Maruyama, Hizawa, Takahashi, and
  Sawada}]{enose2}
\bibinfo{author}{S.~Maruyama}, \bibinfo{author}{T.~Hizawa},
  \bibinfo{author}{K.~Takahashi}, \bibinfo{author}{K.~Sawada},
\newblock \bibinfo{title}{Optical-interferometry-based cmos-mems sensor
  transduced by stress-induced nanomechanical deflection},
\newblock \bibinfo{journal}{Sensors} \bibinfo{volume}{18}
  (\bibinfo{year}{2018}) \bibinfo{pages}{138}.
\bibitem[{Adib et~al.(2018)Adib, Eckstein, Hernandez-Sosa, Sommer, and
  Lemmer}]{enose3}
\bibinfo{author}{M.~Adib}, \bibinfo{author}{R.~Eckstein},
  \bibinfo{author}{G.~Hernandez-Sosa}, \bibinfo{author}{M.~Sommer},
  \bibinfo{author}{U.~Lemmer},
\newblock \bibinfo{title}{Sno$_2$ nanowire-based aerosol jet printed electronic
  nose as fire detector},
\newblock \bibinfo{journal}{IEEE Sens. J.} \bibinfo{volume}{18}
  (\bibinfo{year}{2018}) \bibinfo{pages}{494--500}.
\bibitem[{Nishijima et~al.(2012)Nishijima, Nigorinuma, Rosa, and
  Juodkazis}]{nishijima1}
\bibinfo{author}{Y.~Nishijima}, \bibinfo{author}{H.~Nigorinuma},
  \bibinfo{author}{L.~Rosa}, \bibinfo{author}{S.~Juodkazis},
\newblock \bibinfo{title}{Selective enhancement of infrared absorption with
  metal hole allays},
\newblock \bibinfo{journal}{Opt. Mater. Express} \bibinfo{volume}{2}
  (\bibinfo{year}{2012}) \bibinfo{pages}{1367}.
\bibitem[{Nishijima et~al.(2013)Nishijima, Adachi, Rosa, and
  Juodkazis}]{nishijima2}
\bibinfo{author}{Y.~Nishijima}, \bibinfo{author}{Y.~Adachi},
  \bibinfo{author}{L.~Rosa}, \bibinfo{author}{S.~Juodkazis},
\newblock \bibinfo{title}{Augmented sensitivity of an ir-absorption gas sensor
  employing a metal hole array},
\newblock \bibinfo{journal}{Opt. Mat. Express} \bibinfo{volume}{32}
  (\bibinfo{year}{2013}) \bibinfo{pages}{968}.
\bibitem[{Nishijima et~al.(2017)Nishijima, Suda, Seniutinas, Bal\v{c}ytis, and
  Juodkazis}]{nishijima3}
\bibinfo{author}{Y.~Nishijima}, \bibinfo{author}{S.~Suda},
  \bibinfo{author}{G.~Seniutinas}, \bibinfo{author}{A.~Bal\v{c}ytis},
  \bibinfo{author}{S.~Juodkazis},
\newblock \bibinfo{title}{Plasmonic sensor: towards parts-per-billion level
  sensitivity},
\newblock \bibinfo{journal}{Sens. Mater.} \bibinfo{volume}{29}
  (\bibinfo{year}{2017}) \bibinfo{pages}{1253--1258}.
\bibitem[{Brown et~al.(2015)Brown, Yang, K, Zheng, Nordlander, and
  Halas}]{halas1}
\bibinfo{author}{L.~V. Brown}, \bibinfo{author}{X.~Yang},
  \bibinfo{author}{Z.~K}, \bibinfo{author}{B.~Y. Zheng},
  \bibinfo{author}{P.~Nordlander}, \bibinfo{author}{N.~J. Halas},
\newblock \bibinfo{title}{Fan-shaped gold nanoantennas above reflective
  substrates for surface-enhanced infrared absorption (seira)},
\newblock \bibinfo{journal}{Nano Lett.} \bibinfo{volume}{15}
  (\bibinfo{year}{2015}) \bibinfo{pages}{1272--1280}.
\bibitem[{Dong et~al.(2017)Dong, Yang, Zhang, Cerjan, Zhou, Tseng, Zhang,
  Alabastri, Nordlander, and Halas}]{halas2}
\bibinfo{author}{L.~Dong}, \bibinfo{author}{X.~Yang},
  \bibinfo{author}{C.~Zhang}, \bibinfo{author}{B.~Cerjan},
  \bibinfo{author}{L.~Zhou}, \bibinfo{author}{M.~L. Tseng},
  \bibinfo{author}{Y.~Zhang}, \bibinfo{author}{A.~Alabastri},
  \bibinfo{author}{P.~Nordlander}, \bibinfo{author}{N.~J. Halas},
\newblock \bibinfo{title}{Nanogapped au antennas for ultrasensitive
  surface-enhanced infrared absorption spectroscopy},
\newblock \bibinfo{journal}{Nano Lett.} \bibinfo{volume}{17}
  (\bibinfo{year}{2017}) \bibinfo{pages}{5768--5774}.
\bibitem[{Diem et~al.(2009)Diem, Koschny, and Soukoulis}]{diem}
\bibinfo{author}{M.~Diem}, \bibinfo{author}{T.~Koschny}, \bibinfo{author}{C.~M.
  Soukoulis},
\newblock \bibinfo{title}{Wide-angle perfect absorber/thermal emitter in the
  thz regime},
\newblock \bibinfo{journal}{Phys. Rev. B} \bibinfo{volume}{79}
  (\bibinfo{year}{2009}) \bibinfo{pages}{033103}.
\bibitem[{Miyazaki et~al.(2014)Miyazaki, Kasaya, Iwanaga, Bongseok, Sugimoto,
  and Sakoda}]{miyazaki1}
\bibinfo{author}{H.~Miyazaki}, \bibinfo{author}{T.~Kasaya},
  \bibinfo{author}{M.~Iwanaga}, \bibinfo{author}{C.~Bongseok},
  \bibinfo{author}{Y.~Sugimoto}, \bibinfo{author}{K.~Sakoda},
\newblock \bibinfo{title}{Dual-band infrared metasurface thermal emitter for
  co$_2$ sensing},
\newblock \bibinfo{journal}{Appl. Phys. Lett.} \bibinfo{volume}{105}
  (\bibinfo{year}{2014}) \bibinfo{pages}{121107}.
\bibitem[{Miyazaki et~al.(2008)Miyazaki, Ikeda, Kasaya, Yamamoto, Inoue,
  Fujimura, Kanakugi, Okada, Hatade, and Kitagawa}]{miyazaki2}
\bibinfo{author}{H.~T. Miyazaki}, \bibinfo{author}{K.~Ikeda},
  \bibinfo{author}{T.~Kasaya}, \bibinfo{author}{K.~Yamamoto},
  \bibinfo{author}{Y.~Inoue}, \bibinfo{author}{K.~Fujimura},
  \bibinfo{author}{T.~Kanakugi}, \bibinfo{author}{M.~Okada},
  \bibinfo{author}{K.~Hatade}, \bibinfo{author}{S.~Kitagawa},
\newblock \bibinfo{title}{Thermal emission of two-color polarized infrared
  waves from integrated plasmon cavities},
\newblock \bibinfo{journal}{Appl. Phys. Lett.} \bibinfo{volume}{92}
  (\bibinfo{year}{2008}) \bibinfo{pages}{141114}.
\bibitem[{Ikeda et~al.(2008)Ikeda, Miyazak, Kasaya, Yamamoto, Inoue, Fujimura,
  Kanakugi, Okada, Hatade, and Kitagawa}]{miyazaki3}
\bibinfo{author}{K.~Ikeda}, \bibinfo{author}{H.~T. Miyazak},
  \bibinfo{author}{T.~Kasaya}, \bibinfo{author}{K.~Yamamoto},
  \bibinfo{author}{Y.~Inoue}, \bibinfo{author}{K.~Fujimura},
  \bibinfo{author}{T.~Kanakugi}, \bibinfo{author}{M.~Okada},
  \bibinfo{author}{K.~Hatade}, \bibinfo{author}{S.~Kitagawa},
\newblock \bibinfo{title}{Controlled thermal emission of polarized infrared
  waves from arrayed plasmon nanocavities},
\newblock \bibinfo{journal}{Appl. Phys. Lett.} \bibinfo{volume}{92}
  (\bibinfo{year}{2008}) \bibinfo{pages}{021117}.
\bibitem[{Kusunoki et~al.(2004)Kusunoki, Kohama, Hiroshima, Fukumoto, Takahara,
  and Kobayashi}]{takahara1}
\bibinfo{author}{F.~Kusunoki}, \bibinfo{author}{T.~Kohama},
  \bibinfo{author}{T.~Hiroshima}, \bibinfo{author}{S.~Fukumoto},
  \bibinfo{author}{J.~Takahara}, \bibinfo{author}{T.~Kobayashi},
\newblock \bibinfo{title}{Narrow-band thermal radiation with low rirectivity by
  resonant modes inside tungsten microcavities},
\newblock \bibinfo{journal}{Jpn. J. Appl. Phys.} \bibinfo{volume}{43}
  (\bibinfo{year}{2004}) \bibinfo{pages}{5253}.
\bibitem[{Ueba and Takahara(2012)}]{takahara2}
\bibinfo{author}{Y.~Ueba}, \bibinfo{author}{J.~Takahara},
\newblock \bibinfo{title}{Spectral control of thermal radiation by metasurface
  with split-ring resonator},
\newblock \bibinfo{journal}{Appl. Phys. Express} \bibinfo{volume}{5}
  (\bibinfo{year}{2012}) \bibinfo{pages}{122001}.
\bibitem[{Maruyama et~al.(2001)Maruyama, Kashiwa, Yugami, and
  Esashi}]{maruyama}
\bibinfo{author}{S.~Maruyama}, \bibinfo{author}{T.~Kashiwa},
  \bibinfo{author}{H.~Yugami}, \bibinfo{author}{M.~Esashi},
\newblock \bibinfo{title}{Thermal radiation from two-dimensionally confined
  modes in microcavities},
\newblock \bibinfo{journal}{Appl. Phys. Lett.} \bibinfo{volume}{79}
  (\bibinfo{year}{2001}) \bibinfo{pages}{1393}.
\bibitem[{Liu et~al.(2010)Liu, Mesch, Weiss, Hentschel, and
  Giessen}]{ppasensor}
\bibinfo{author}{N.~Liu}, \bibinfo{author}{M.~Mesch},
  \bibinfo{author}{T.~Weiss}, \bibinfo{author}{M.~Hentschel},
  \bibinfo{author}{H.~Giessen},
\newblock \bibinfo{title}{Infrared perfect absorber and its application as
  plasmonic sensor},
\newblock \bibinfo{journal}{Nano Lett.} \bibinfo{volume}{10}
  (\bibinfo{year}{2010}) \bibinfo{pages}{2342--2348}.
\bibitem[{Hedayati et~al.(2014)Hedayati, Faupel, and Elbahri}]{ppareview}
\bibinfo{author}{M.~K. Hedayati}, \bibinfo{author}{F.~Faupel},
  \bibinfo{author}{M.~Elbahri},
\newblock \bibinfo{title}{Review of plasmonic nanocomposite metamaterial
  absorber},
\newblock \bibinfo{journal}{Materials} \bibinfo{volume}{7}
  (\bibinfo{year}{2014}) \bibinfo{pages}{1221--1248}.
\bibitem[{DESHPANDE et~al.(2017)DESHPANDE, PORS, and Bozhevolnyi}]{sergey}
\bibinfo{author}{R.~DESHPANDE}, \bibinfo{author}{A.~PORS},
  \bibinfo{author}{S.~I. Bozhevolnyi},
\newblock \bibinfo{title}{Third-order gap plasmon based metasurfaces for
  visible light},
\newblock \bibinfo{journal}{Opt. Express} \bibinfo{volume}{25}
  (\bibinfo{year}{2017}) \bibinfo{pages}{12508--12517}.
\bibitem[{Leveque and Martin(2006)}]{oliver}
\bibinfo{author}{G.~Leveque}, \bibinfo{author}{O.~J.~F. Martin},
\newblock \bibinfo{title}{Optical interactions in a plasmonic particle coupled
  to a metallic film},
\newblock \bibinfo{journal}{Opt. Express} \bibinfo{volume}{14}
  (\bibinfo{year}{2006}) \bibinfo{pages}{9971--9981}.
\bibitem[{Nishijima et~al.(2017)Nishijima, Bal\v{c}ytis, Seniutinas, Juodkazis,
  Arakawa, Okazaki, and Petruskevicius}]{nishijima4}
\bibinfo{author}{Y.~Nishijima}, \bibinfo{author}{A.~Bal\v{c}ytis},
  \bibinfo{author}{G.~Seniutinas}, \bibinfo{author}{S.~Juodkazis},
  \bibinfo{author}{T.~Arakawa}, \bibinfo{author}{S.~Okazaki},
  \bibinfo{author}{R.~Petruskevicius},
\newblock \bibinfo{title}{Plasmonic hydrogen sensor at infrared wavelength},
\newblock \bibinfo{journal}{Sens. Mater.} \bibinfo{volume}{29}
  (\bibinfo{year}{2017}) \bibinfo{pages}{1269--1274}.
\bibitem[{Nishijima et~al.(????)Nishijima, Bal\v{c}ytis, Naganuma, Seniutinas,
  and Juodkazis}]{PPAs}
\bibinfo{author}{Y.~Nishijima}, \bibinfo{author}{A.~Bal\v{c}ytis},
  \bibinfo{author}{S.~Naganuma}, \bibinfo{author}{G.~Seniutinas},
  \bibinfo{author}{S.~Juodkazis},
\newblock \bibinfo{title}{Tailoring metal and insulator contributions in
  plasmonic perfect absorber metasurfaces},
\newblock \bibinfo{journal}{submitted to ChemRxiv}  (????).
\bibitem[{Pu et~al.(2011)Pu, Hu, Wang, Huang, Zhao, Wang, Feng, and Luo}]{ppa}
\bibinfo{author}{M.~Pu}, \bibinfo{author}{C.~Hu}, \bibinfo{author}{M.~Wang},
  \bibinfo{author}{C.~Huang}, \bibinfo{author}{Z.~Zhao},
  \bibinfo{author}{C.~Wang}, \bibinfo{author}{Q.~Feng},
  \bibinfo{author}{X.~Luo},
\newblock \bibinfo{title}{Design principles for infrared wide-angle perfect
  absorber based on plasmonic structure},
\newblock \bibinfo{journal}{Opt. Express} \bibinfo{volume}{19}
  (\bibinfo{year}{2011}) \bibinfo{pages}{17413--17420}.
\bibitem[{Brewster(1992)}]{rad}
\bibinfo{author}{M.~Q. Brewster},
\newblock \bibinfo{title}{Thermal radiative transfer and properties}
  (\bibinfo{year}{1992}).
\bibitem[{Komatsua et~al.(2015)Komatsua, Bal\v{c}ytis, Seniutinas, Yamamura,
  Nishijima, and Juodkazis}]{komatsu}
\bibinfo{author}{R.~Komatsua}, \bibinfo{author}{A.~Bal\v{c}ytis},
  \bibinfo{author}{G.~Seniutinas}, \bibinfo{author}{T.~Yamamura},
  \bibinfo{author}{Y.~Nishijima}, \bibinfo{author}{S.~Juodkazis},
\newblock \bibinfo{title}{Plasmonic photo-thermoelectric energy converter with
  black-si absorber},
\newblock \bibinfo{journal}{Sol. Ener. Mater. Sol. Cell} \bibinfo{volume}{143}
  (\bibinfo{year}{2015}) \bibinfo{pages}{72--77}.
\bibitem[{Greffet et~al.(2002)Greffet, Carminati, Joulain, Mulet, Mainguy, and
  Chen}]{nature}
\bibinfo{author}{J.-J. Greffet}, \bibinfo{author}{R.~Carminati},
  \bibinfo{author}{K.~Joulain}, \bibinfo{author}{J.-P. Mulet},
  \bibinfo{author}{S.~Mainguy}, \bibinfo{author}{Y.~Chen},
\newblock \bibinfo{title}{Coherent emission of light by thermal sources},
\newblock \bibinfo{journal}{Nature} \bibinfo{volume}{416}
  (\bibinfo{year}{2002}) \bibinfo{pages}{61--64}.
\bibitem[{Liu et~al.(2015)Liu, Guler, Lagutchev, Kildishev, Malis, Boltasseva,
  and Shalaev}]{liu}
\bibinfo{author}{J.~Liu}, \bibinfo{author}{U.~Guler},
  \bibinfo{author}{A.~Lagutchev}, \bibinfo{author}{A.~Kildishev},
  \bibinfo{author}{O.~Malis}, \bibinfo{author}{A.~Boltasseva},
  \bibinfo{author}{V.~M. Shalaev},
\newblock \bibinfo{title}{Quasi-coherent thermal emitter based on refractory
  plasmonic materials},
\newblock \bibinfo{journal}{Opt. Mater. Express} \bibinfo{volume}{5}
  (\bibinfo{year}{2015}) \bibinfo{pages}{2721--2728}.

\end{thebibliography}







\end{document}